\documentclass[preprint]{aastex}
\usepackage{amsmath, amsfonts, amssymb, natbib}

\begin{document}

\title{Galaxies M32 and NGC 5102 Confirm a Near-infrared Spectroscopic Chronometer}

\author{Jesse Miner\altaffilmark{1,2}, James A. Rose\altaffilmark{1}, Gerald Cecil\altaffilmark{1,2}}

\altaffiltext{1}{Department of Physics and Astronomy, University of North Carolina, Chapel Hill, NC 27516, USA}

\altaffiltext{2}{Visiting Astronomer at the Infrared Telescope Facility, which is operated by the University of Hawaii under Cooperative Agreement no.\ NNX-08AE38A with the National Aeronautics and Space Administration, Science Mission Directorate, Planetary Astronomy Program.}

\begin{abstract}

We present near infrared (NIR) IRTF/SpeX spectra of the intermediate-age
galaxy M32 and the post-starburst galaxy NGC 5102. We show that features
from thermally-pulsing asymptotic giant branch (TP-AGB) and main
sequence turn-off (MSTO) stars yield similar ages to those derived from
optical spectra.
The TP-AGB can dominate the NIR flux of a coeval stellar population between
$\sim$0.1 and $\sim$2~Gyr, and the strong features of (especially
C-rich) TP-AGB stars are useful chronometers in integrated light studies.
Likewise, the Paschen series in MSTO stars is stongly dependent on age
and is an indicator of a young stellar component in integrated spectra. 
We define
four NIR spectroscopic indices to measure the strength of 
absorption features from both C-rich TP-AGB stars and hydrogen 
features in main sequence stars, in a preliminary effort to construct
a robust chronometer that probes the contributions from stars in 
different evolutionary phases.
By comparing the values of the indices measured in M32 and NGC~5102 
to those in the \citet{m05} stellar
population synthesis models for various ages and metallicities,
we show that model predictions for the ages of the nuclei of M32 and NGC~5102 
agree with previous results obtained from
integrated optical spectroscopy and CMD analysis of the giant
branches. The indices discriminate between 
an intermediate age population of $\sim$3--4~Gyr, a younger 
population of $\lesssim$1~Gyr, and can also detect the signatures of very
young $\lesssim$100~Myr populations. 
%This technique will be useful
%for studying the formation history of galaxies in the early
%universe with the \emph{James Webb Space Telescope}.

\end{abstract}


\section{Introduction}\label{sec:int}

Integrated spectroscopy of galaxies has been performed primarily in the 
optical because much of the (blue) optical light comes
from the well-understood main sequence turnoff stars. As one pushes into
the near infrared (NIR), the integrated light of a stellar population
is dominated by very luminous stars in later stages of stellar evolution,
such as thermally pulsing asymptotic giant branch (TP-AGB) stars.
Beginning with \citet{renzini81} and
continuing with, e.g.,\ \citet{frogel90}, \citet{bressan98}, 
\citet{m98}, \citet{lancon99}, 
\citet{lancon02}, it has become clear
that the unique features and high luminosity of TP-AGB stars
profoundly affect the NIR integrated light of stellar populations.
Specifically, for stellar populations
with ages $\sim$100~Myr to $\sim$1~Gyr, the TP-AGB can
contribute up to $\sim$60\% of the K-band flux \citep[depending
on metallicity, see][]{m05}. Strong molecular features, from CN,
CO, C$_2$, and H$_2$O, in TP-AGB stellar atmospheres should appear in
the integrated NIR spectrum of populations in this age range. 

Unfortunately, relatively poorly
understood processes such as mass loss, 
convective transport of heavy elements to the surface, 
and thermal pulsation influence the late stages of stellar
evolution and thus predictions for the integrated spectrum.
Furthermore, the ratio of C-rich to O-rich AGB stars
in a population is known to depend strongly on metallicity,
meaning that the contribution from TP-AGB stars can vary
drastically not only with age but also composition \citep{mouhcine03}.
However, observational work has constrained
these effects on the integrated light of stellar populations
with observations of Magellanic Cloud and Galactic clusters, 
and the predictive potential
of NIR spectroscopic features has been demonstrated 
\citep[e.g.,\ ][]{lancon99,mouhcine02,m05}. 
In short, TP-AGB features should be traceable in galaxies 
whose NIR light is dominated by
stars in the $\sim$100~Myr to $\sim$1~Gyr age range.  

For example, \citet{mouhcine02a} have detected the presence of 
TP-AGB stars in the NIR spectrum of a young star cluster 
in the galaxy NGC 7252, and have verified that the 
predicted features expected in a young population ($\sim$300~Myr) are 
present in the NIR spectrum.
This successful detection of TP-AGB features in NGC~7252 thus motivates 
additional NIR spectroscopy of bright, nearby galaxies with young 
populations to further test the use of NIR spectral indicators as 
chronometers.
%This successful detection of TP-AGB features
%in integrated light has inspired a search for other bright, nearby
%galaxies with young populations where high signal-to-noise (SNR) ratio
%NIR spectroscopy can provide measurements of individual features. 
We also note that while light from the TP-AGB can dominate the NIR flux,
the main sequence turn-off (MSTO) also contributes an appreciable
amount to the integrated light, and hydrogen features (specifically
the Paschen series) due to MSTO stars will be visible. 
To investigate the behavior of the TP-AGB and MSTO features as a
function of both age and metallicity we consulted the 
\citet[][hereafter M05]{m05} stellar population synthesis models, 
which include a careful treatment of the TP-AGB.
In this Letter we use NIR spectra of two galaxies with very different
star formation histories to test if population
synthesis models of TP-AGB and MSTO features can differentiate
between young and intermediate-age populations.

%By chosing
%two test galaxies with very different star formation histories, we
%investigate the ability for TP-AGB and MSTO features to differentiate
%between young and intermediate-age populations, and test the predictive
%accuracy of the M05 models.***

%In this Letter we 
%illustrate this point through high signal-to-noise ratio (SNR) NIR 
%spectroscopy of two nearby galaxies.

%Many spectroscopic studies have targeted the compact elliptical 
%galaxy M32, 
Because its high nuclear surface brightness and proximity 
allow for both high
SNR spectroscopy and resolved stellar photometry
of the giant branch, the intermediate age compact elliptical galaxy 
M32 has been extensively studied. Recent work
has included integrated optical spectroscopy of
the nucleus and extranuclear region out to the half-light
radius $R_e$ \citep{delburgo01,worthey04,rose05,coelho09}, 
and NIR imaging of the giant branches
\citep{davidge07}. These studies find that the nuclear light is
dominated by an approximately solar metallicity
intermediate-age population ($\sim$3 -- 4~Gyr). 
Because its stellar content has been constrained
by several established age-dating techniques, M32 is 
ideal for testing other predictive techniques. 

Likewise, the blue S0 galaxy NGC~5102 is close and bright enough
for high SNR integrated spectroscopy plus
resolved CMD analysis of its brightest stars. It
is classified as a post-starburst (PSB) galaxy because its
strong Balmer-line absorption and absence of emission
indicates that star formation terminated recently.
Studies by, e.g., \citet{deharveng97,davidge08} have indicated
that the nuclear region of NGC 5102 has undergone a termination
of star formation within the past $\sim$10--100~Myr, after a several
hundred Myr period of activity, guaranteeing that there
is an appreciable young population with a mean age of a few 100 Myr.
%The derived star formation history (SFH) of the nuclear
%region of NGC~5102 includes 
%activity within the past several hundred Myr, terminating
%$\sim$10 -- $\sim$100~Myr ago \citep{deharveng97,davidge08}, and 
%the emitted light will be dominated by 
The intermediate
age stars in M32 and the younger stars in NGC~5102 are
a useful pairing because most of the NIR
light emitted by these two galaxies comes from stars in
different evolutionary stages: the younger population will
be dominated by the TP-AGB and a blue MSTO, while in the intermediate
age population most of the light comes from the red giant branch
and an older MSTO (see Fig.~13 in M05).

In this Letter we
define four age-sensitive NIR spectroscopic indices, two of which
measure previously defined TP-AGB features, and two that measure 
Paschen series absorption lines, and compare the 
observed values in M32 and NGC~5102 to predictions of the M05
stellar population models. We demonstrate that the derived ages from the
NIR indices agree closely with previously published ages determined from 
optical spectroscopy, and thus provide evidence to support the
reliability of NIR spectral signatures as a useful chronometer for
galaxy evolution.

\section{Observations and Stellar Population Models}

On the nights of 12 and 13 May 2010, we observed NGC~5102 with
the SpeX spectrograph on the NASA Infrared 
Telescope Facility \citep{rayner03}. The detector is a
1024$\times$1024 Aladdin~3 InSb array. We used the short 
cross-dispersed (SXD) mode, with a 0$\farcs$8 slit
(spectral $R=750$) aligned with the parallactic angle. We took
two ABBA object/sky sequences for a total of
eight on-object exposures of 120~s each.
The SXD mode simultaneously covers $\lambda\lambda0.81-2.4$~$\mu$m, 
so avoids issues such as order overlap and
variable sky features that plague single-order infrared
spectroscopy. Standard SpeX calibrations (arc lamps and
flats) were performed after the exposures, and we observed
a standard A0~V star at similar airmass to remove
atmospheric absorption and calibrate fluxes. 
The nuclear region of M32 was observed on 10 June 2010 
with the same instrumental 
setup, number of exposures and integration time. 

\begin{figure*}
\begin{center}
\includegraphics[scale=0.6]{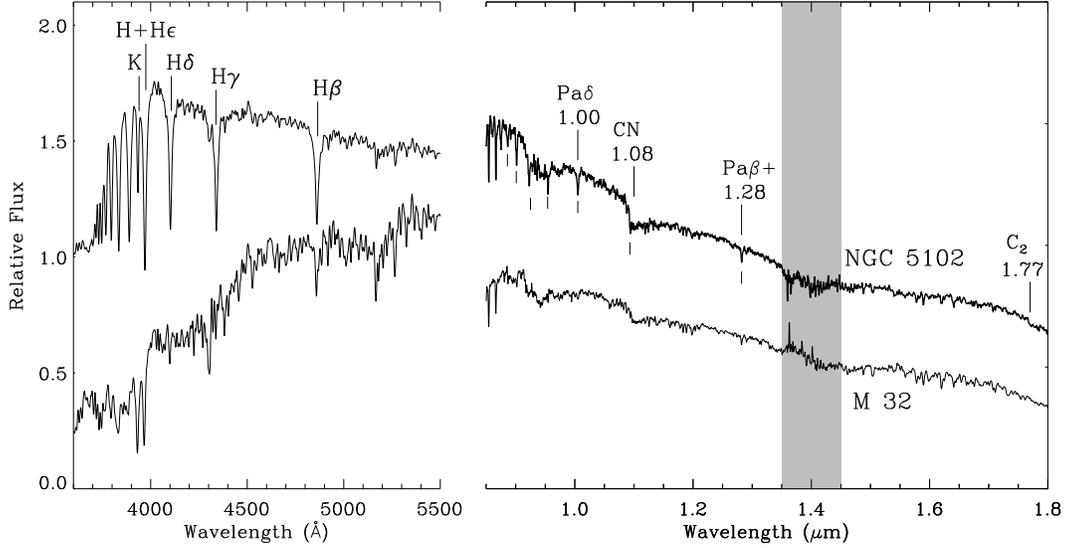}
\caption{\small Optical (for reference) and NIR spectra of NGC~5102 
and M32 are plotted.
The optical spectrum of NGC~5102 is from the Goodman High Throughput
Spectrograph \citep{clemens04} on the 4.1~m SOAR telescope at Cerro
Pachon, Chile, while the optical spectrum of M32 was taken
with the FOCAS spectrograph on the 8~m Subaru telescope at
Mauna Kea, Hawaii \citep[see][]{rose05}. The NIR spectra were 
taken with the 
SpeX spectrograph for this paper.  The region of low 
atmospheric transparency is shaded. The J and H bands are shown, 
as we have defined no indices in the K band. The Paschen series is marked
with the lower ticks to illustrate the differences between the spectra.
Note the stronger Balmer lines in
the optical spectrum of NGC~5102 than in M32, which is also reflected in
the stronger Paschen series in the NIR spectrum of NGC~5102.
\label{fig:spec}}
\end{center}
\end{figure*}

Spectral extraction and atmospheric correction were performed
using the Spextool package, version~3.4 \citep{cushing04}.
See \citet{vacca03} for a detailed description of
the telluric correction process.  
The final spectra of six separate orders
were coadded and combined, and are displayed in Figure~\ref{fig:spec},
along with optical spectra.

\begin{deluxetable}{l l l c c c c c c}
  \tabletypesize{\scriptsize}  
  \tablecaption{Index Definitions and Values\label{tab:feat}} 
  \tablewidth{0pt} 
  \tablehead{
    \colhead{Feature} & 
    \colhead{Band 1 ($\mu$m)} &
    \colhead{Band 2 ($\mu$m)} & 
    \colhead{Line ($\mu$m)} &
    \colhead{M32} &
    \colhead{NGC~5102}
  }
  \startdata
  
  $1.00\mu$m Pa$\delta$ & 0.997--1.000 & 1.012--1.015 & 1.002--1.011 & 0.84 \AA & 2.63 \AA \\
  $1.08\mu$m CN & 1.060--1.070 & 1.095--1.105 & -- & 1.10 & 1.20 \\
  $1.28\mu$m Pa$\beta+$ & 1.258--1.265 & 1.297--1.305 & 1.275--1.289 & 1.75 \AA & 2.89 \AA \\
  $1.77\mu$m C$_2$& 1.753--1.763 & 1.775--1.785 & -- & 1.06 & 1.11

  \enddata
  \tablecomments{Bandpasses used to define spectral indices. The
1.00 and 1.28 $\mu$m Paschen features are equivalent widths, while the two 
other indices are flux ratios of narrow bands bracketing molecular 
absorption breaks due to TP-AGB stars. The index values for M32 and NGC~5102
are included.}
\end{deluxetable}

\begin{figure}
\begin{center}
\includegraphics[scale=0.5]{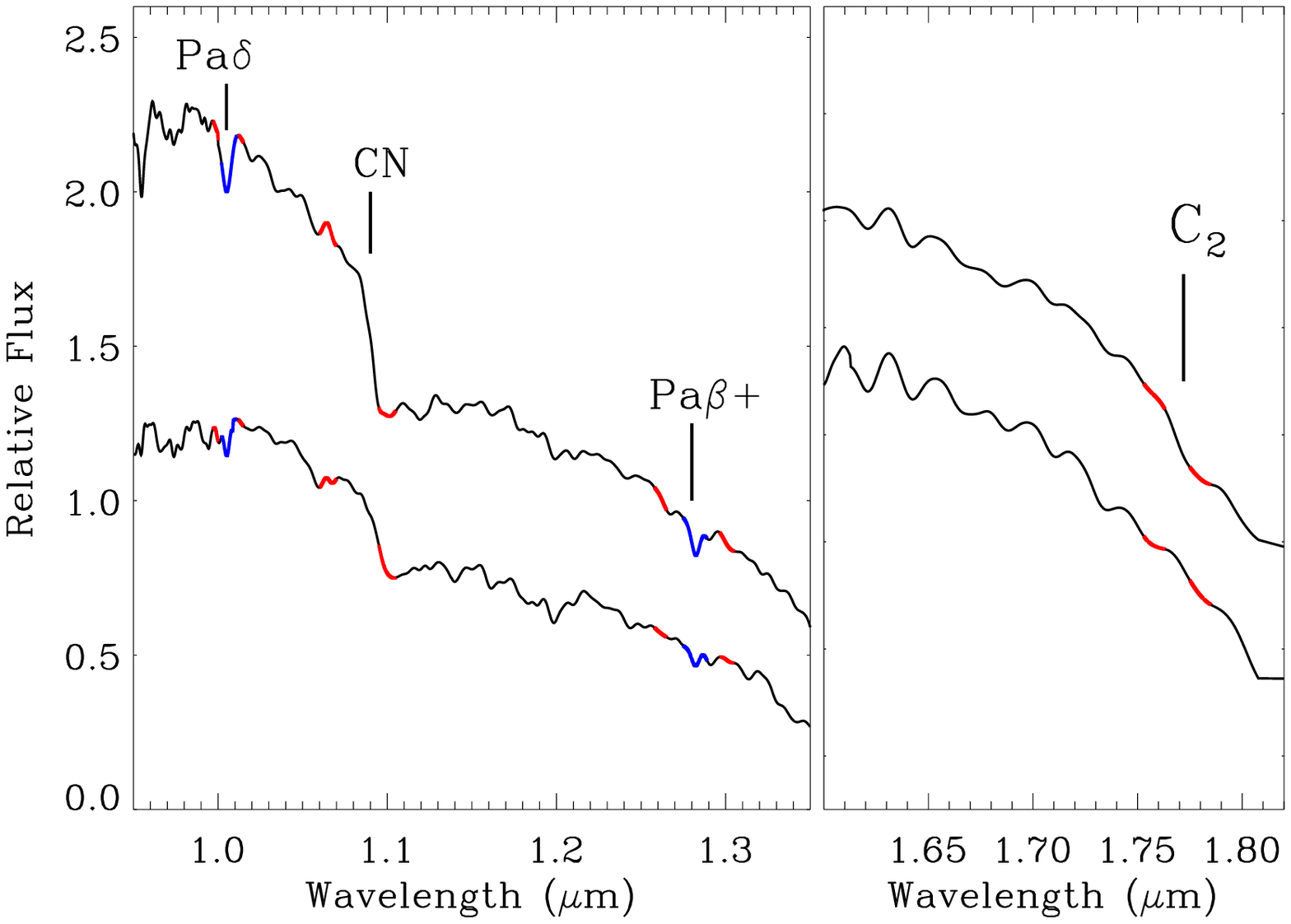}
\caption{\small Wavelength regions of spectral indices for NGC~5102 (top)
and M32 (bottom) smoothed to the resolution of the M05 models, 
with flux bands shown in red, and line regions in blue. 
\label{fig:feat}}
\end{center}
\end{figure}

We define four spectroscopic indices (see Table~\ref{tab:feat}): two
are flux ratios of narrow bands bracketing absorption
breaks (1.08$\mu$m CN, 1.77$\mu$m C$_2$), and two
are equivalent widths of hydrogen Paschen lines (1.00$\mu$m Pa$\delta$, 
1.28$\mu$m Pa$\beta+$), where the ``$+$'' reminds the reader that
due to the low resolution of the
models, other lines are present in the Pa$\beta$ equivalent width feature.
The absorption breaks arise from the cool, luminous stars on the TP-AGB,
while the Paschen lines are contributed by hot main sequence turnoff stars
in young populations, and thus this method probes two independent
stellar phases with very different physical origins (the cool, extended
atmospheres of TP-AGB stars vs. the hot atmospheres of MSTO stars).  
An enhanced view of the spectra of M32 and NGC~5102 in the regions of
the four spectral indices is shown in Figure~\ref{fig:feat}.
To track the dependence of these four indices on age and metallicity we
have measured them in the M05 simple stellar population (SSP) models,
which cover a large range in both age and metal abundance, and provide an
output integrated spectrum for each age and metallicity. In all cases we
have used a Kroupa initial mass function and a red horizontal branch.

\section{Results}

\begin{figure*}
\begin{center}
\includegraphics[scale=0.7]{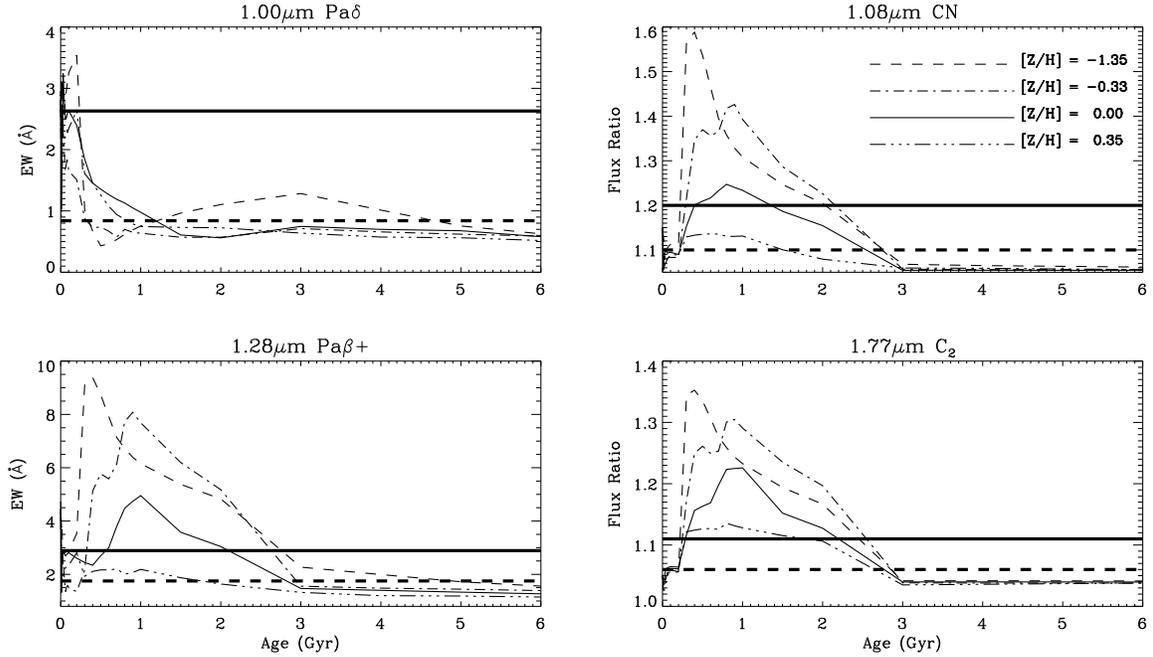}
\caption{\small NIR spectral indices derived from the M05 
models are plotted as 
a function of age, for four metal abundances.  Thick horizontal lines 
denote the observed values for NGC~5102 (solid line) and for M32 (dashed 
line), respectively. The formal (photon statistical) error bars in the 
NGC~5102 and M32 indices are smaller than the thickness of the plotted 
lines.  We note that the Pa$\beta+$ index is contaminated with other 
features.
\label{fig:inds}}
\end{center}
\end{figure*}

As can be seen qualitatively in Figure~\ref{fig:spec}, there are clear
differences between both the optical and NIR spectra of the recent PSB galaxy
NGC~5102 and the intermediate-age M32.  In the optical the Balmer lines are
much more prominent, and the \ion{Ca}{2} K line weaker, 
in NGC~5102 than in
M32.  Likewise, the Paschen line series and the two break features
due to the TP-AGB are significantly stronger in the NIR spectrum of 
NGC~5102 than M32. To provide a quantitative foundation to
these differences and to make a direct comparison with the predictions for
spectral index behavior of the M05 models, in Figure~\ref{fig:inds}
values for the four above-defined indices are plotted as a function of age
for the M05 models, at four different metallicities, along with the
observed values for NGC~5102 and M32.
Due to the high SNR of the observed spectra, the
formal (photon statistical) error bars for the NGC~5102 and M32 index values
are actually smaller than the width of the plotted lines.  Systematic errors 
due to, e.g., flat-fielding issues, subtraction of night sky emission lines,
and removal of telluric features, clearly dominate the true uncertainties,
and await a more detailed analysis.

It is readily seen that the two absorption break indices,
as well as the Pa$\beta+$ index, first increase 
with increasing age as the TP-AGB becomes
more prominent, and then decline at ages greater than $\sim$1~Gyr,
as the RGB overtakes the TP-AGB as the dominant contributor in the NIR. 
The Pa$\delta$ index behaves similarly, but with the index first rising and 
then falling more rapidly with age (at ~100 Myr).  
The two galaxies are seen to have quite distinct SFHs; the large break
indices and large Paschen index values in NGC~5102 
argue for a young ($\lesssim$2~Gyr) population with appreciable
flux from the TP-AGB, and from upper main sequence stars,
while M32 shows little contribution from the TP-AGB or young stars.

Because both NGC~5102 and M32 must have had complicated SFHs and chemical
enrichment histories, it is certainly an oversimplification to compare
their integrated spectra to SSP model predictions.  The derived ages are
really an SSP-equivalent age \citep{serra07}, in that the mean age
derived from the indices is weighted by the contributions from the entire
SFH of the galaxy, which tends to more heavily weight the younger, more
luminous, populations.  Nevertheless, we can 
demonstrate that inferences about the SFHs of these two 
galaxies from the NIR spectra
concur with previous results obtained at optical wavelengths.  In the case
of M32, the low values of all four NIR indices consistently argue for an
SSP-equivalent age $\gtrsim$3~Gyr, which is in accord with the ages of
$\sim$4~Gyr obtained from optical spectra by Worthey (2004) and \citet{rose05}
for the central region; \citet{davidge07} also find an 
intermediate-age population in M32 based on resolved star photometry of the
RGB.  \citet{coelho09} find evidence for an old ($>$10~Gyr) population
underlying the intermediate age population, each contributing roughly half
of the optical light.  
The implication is that the SSP-equivalent age results
from the combination of an old population, and one no older than
$\sim$4~Gyr.  In principle, for instance, the SSP-equivalent age of 4~Gyr
could arise from the combination of a young, $\sim$1~Gyr, population
superposed on the old population.  However, in \citet{rose94} it was found
on the basis of spectral indices in the blue that any contribution from a
1 Gyr population must be very small.  The NIR spectral indices for M32
plotted in Figure~\ref{fig:inds} certainly supports the conclusion from the
optical indices that any contribution from a population as young as 1~Gyr
must be small.

Turning now to NGC~5102, the most striking aspect of Figure~\ref{fig:inds}
is the contrast between the consistently higher NIR indices in NGC~5102 
and those in M32, implying that while the light of M32 
is dominated by an
intermediate-age population, that of NGC~5102 is dominated by a younger 
population in which both the TP-AGB and hot young main sequence stars make
a strong contribution.  The TP-AGB signature is evident in the high 
observed values of the 1.08$\mu$m and 1.77$\mu$m break features, while 
the young main sequence contribution is evident in the high value of the 
Pa$\delta$ index at 1.00~$\mu$m and to a lesser extent the 
partially-contaminated Pa$\beta+$ index at 1.28~$\mu$m.

Two aspects of the behavior of the NIR indices with age and
metallicity complicate their interpretation.
First, as mentioned,
the break indices initially rise with age at $\sim$100 Myr as the
TP-AGB phase develops, then decline
beyond $\sim$1~Gyr as the RGB develops.  The Paschen indices behave
similarly as the hydrogen lines in upper main
sequence stars first strengthen and then weaken with the turnoff 
temperature.
Thus two age solutions are possible  for many index values.
However, the Pa$\delta$ index peaks at substantially
younger ages than the other three indices, so can often lift the age 
degeneracy.
Second, the timing and duration of the index peaks depend on metallicity.
This is due in part to the strong metallicity dependence of the ratio
of C-rich to O-rich AGB stars in a population: metal-poor populations will
have a larger fraction of C-rich giants than their metal-rich counterparts.
For NGC~5102 we can assume approximately solar metallicity,
because \citet{davidge08} finds that the metallicity distribution
function (MDF) of its disk RGB stars
peaks at [Fe/H]$\sim$-0.1, whereas \citet{beaulieu10} report that
the young nuclear population has [Fe/H]$=0.0$.
The solar metallicity models in Figure~\ref{fig:inds}
show that each of the observed indices for NGC 5102 is consistent with
an age of $\sim$300~Myr. The shape of the
absorption breaks and the 1.28$\mu$m indices yield
two possible ages from any measurement, but
the 1.00$\mu$m Pa$\delta$ feature is sufficiently distinct in age
sensitivity to distinguish between a younger and more intermediate age
population, and constrains the age to $\sim$300~Myr.
As mentioned, recent studies of NGC~5102
report a period of star formation lasting several hundred Myr
and ending within the past $\sim$100~Myr, showing
that our index values are consistent
with the results from established age-dating techniques.

Our measured NIR spectral indices in M32 and NGC~5102 reinforce 
the importance of TP-AGB constraints on galaxy ages obtained from NIR
spectra,
especially when combined with features that probe the MSTO.
This technique is particularly useful for distinguishing intermediate age 
($\sim$3-5~Gyr) galaxies from those dominated by a young ($\sim$1~Gyr)
population.  We can further discriminate for the presence of a very
young ($\sim$100~Myr) population on the basis of the Pa$\delta$ index.
Hence our analysis demonstrates that NIR indices allow for a fairly
complete description of age and metallicity for a range of stellar
population ages that will become accessible with the large lookback times
observable with the next generation IR space telescope, the \emph{James Webb 
Space Telescope}. Specifically,
this chronometer will aid in testing models of galaxy
formation in the early universe: as a large number of high redshift 
galaxies are observed in the mid-IR, formation timescales can 
be constrained with the (restframe) NIR indices presented here.
Because $z>2$ galaxies are $\la$2~Gyr old (assuming standard $\Lambda$-CDM
cosmology), the TP-AGB and young main sequence turnoff will provide much of the 
integrated flux, and the time of their formation
can be determined from the measured NIR indices.

\section{Conclusion}

We present an analysis of integrated NIR SpeX SXD spectra of the 
nuclear regions of M32 and NGC~5102, and show that mean ages 
determined from spectroscopic indices agree with previous
studies, with the important result that this method can differentiate
between young and intermediate-age populations. The indices probe
contributions from two different stellar evolutionary phases,
the TP-AGB and the MSTO, which effectively provides two independent
chronometers. Galaxy ages are derived by comparing index values to 
those measured in M05 model SSPs, indicating an accurate 
treatment of the TP-AGB in the models. This method of defining NIR indices
is the first step towards a robust NIR spectroscopic age-dating technique,
which will be particularly useful
when applied to high redshift spectroscopic surveys undertaken by
the next generation of infrared observatories.

\acknowledgements
We thank Claudia Maraston for enlightening discussions
and assistance with the SPS models. JM and JAR acknowledge support
from NC Space Grant programs.

{\it Facilities:} \facility{IRTF (SpeX)}


\begin{thebibliography}{20}
\expandafter\ifx\csname natexlab\endcsname\relax\def\natexlab#1{#1}\fi

\bibitem[{{Beaulieu} {et~al.}(2010){Beaulieu}, {Freeman}, {Hidalgo}, {Norman},
  \& {Quinn}}]{beaulieu10}
{Beaulieu}, S.~F., {Freeman}, K.~C., {Hidalgo}, S.~L., {Norman}, C.~A., \&
  {Quinn}, P.~J. 2010, \aj, 139, 984

\bibitem[{{Bressan} {et~al.}(1998){Bressan}, {Granato}, \& {Silva}}]{bressan98}
{Bressan}, A., {Granato}, G.~L., \& {Silva}, L. 1998, \aap, 332, 135

\bibitem[{{Clemens} {et~al.}(2004){Clemens}, {Crain}, \&
  {Anderson}}]{clemens04}
{Clemens}, J.~C., {Crain}, J.~A., \& {Anderson}, R. 2004, in Presented at the
  Society of Photo-Optical Instrumentation Engineers (SPIE) Conference, Vol.
  5492, Society of Photo-Optical Instrumentation Engineers (SPIE) Conference
  Series, ed. {A.~F.~M.~Moorwood \& M.~Iye}, 331--340

\bibitem[{{Coelho} {et~al.}(2009){Coelho}, {Mendes de Oliveira}, \& {Cid
  Fernandes}}]{coelho09}
{Coelho}, P., {Mendes de Oliveira}, C., \& {Cid Fernandes}, R. 2009, \mnras,
  396, 624

\bibitem[{{Cushing} {et~al.}(2004){Cushing}, {Vacca}, \& {Rayner}}]{cushing04}
{Cushing}, M.~C., {Vacca}, W.~D., \& {Rayner}, J.~T. 2004, \pasp, 116, 362

\bibitem[{{Davidge}(2008)}]{davidge08}
{Davidge}, T.~J. 2008, \aj, 135, 1636

\bibitem[{{Davidge} \& {Jensen}(2007)}]{davidge07}
{Davidge}, T.~J., \& {Jensen}, J.~B. 2007, \aj, 133, 576

\bibitem[{{Deharveng} {et~al.}(1997){Deharveng}, {Jedrzejewski}, {Crane},
  {Disney}, \& {Rocca-Volmerange}}]{deharveng97}
{Deharveng}, J., {Jedrzejewski}, R., {Crane}, P., {Disney}, M.~J., \&
  {Rocca-Volmerange}, B. 1997, \aap, 326, 528

\bibitem[{{del Burgo} {et~al.}(2001){del Burgo}, {Peletier}, {Vazdekis},
  {Arribas}, \& {Mediavilla}}]{delburgo01}
{del Burgo}, C., {Peletier}, R.~F., {Vazdekis}, A., {Arribas}, S., \&
  {Mediavilla}, E. 2001, \mnras, 321, 227

\bibitem[{{Frogel} {et~al.}(1990){Frogel}, {Mould}, \& {Blanco}}]{frogel90}
{Frogel}, J.~A., {Mould}, J., \& {Blanco}, V.~M. 1990, \apj, 352, 96

\bibitem[{{Lan{\c c}on} \& {Mouhcine}(2002)}]{lancon02}
{Lan{\c c}on}, A., \& {Mouhcine}, M. 2002, \aap, 393, 167

\bibitem[{{Lan{\c c}on} {et~al.}(1999){Lan{\c c}on}, {Mouhcine}, {Fioc}, \&
  {Silva}}]{lancon99}
{Lan{\c c}on}, A., {Mouhcine}, M., {Fioc}, M., \& {Silva}, D. 1999, \aap, 344,
  L21

\bibitem[{{Maraston}(1998)}]{m98}
{Maraston}, C. 1998, \mnras, 300, 872

\bibitem[{{Maraston}(2005)}]{m05}
---. 2005, \mnras, 362, 799

\bibitem[{{M{\'a}rmol-Queralt{\'o}} {et~al.}(2009){M{\'a}rmol-Queralt{\'o}},
  {Cardiel}, {S{\'a}nchez-Bl{\'a}zquez}, {Trager}, {Peletier}, {Kuntschner},
  {Silva}, {Cenarro}, {Vazdekis}, \& {Gorgas}}]{marmol09}
{M{\'a}rmol-Queralt{\'o}}, E., {Cardiel}, N., {S{\'a}nchez-Bl{\'a}zquez}, P.,
  {Trager}, S.~C., {Peletier}, R.~F., {Kuntschner}, H., {Silva}, D.~R.,
  {Cenarro}, A.~J., {Vazdekis}, A., \& {Gorgas}, J. 2009, \apjl, 705, L199

\bibitem[{{Mouhcine} \& {Lan{\c c}on}(2002)}]{mouhcine02}
{Mouhcine}, M., \& {Lan{\c c}on}, A. 2002, \aap, 393, 149

\bibitem[{{Mouhcine} \& {Lan{\c c}on}(2003)}]{mouhcine03}
---. 2003, \mnras, 338, 572

\bibitem[{{Mouhcine} {et~al.}(2002){Mouhcine}, {Lan{\c c}on}, {Leitherer},
  {Silva}, \& {Groenewegen}}]{mouhcine02a}
{Mouhcine}, M., {Lan{\c c}on}, A., {Leitherer}, C., {Silva}, D., \&
  {Groenewegen}, M.~A.~T. 2002, \aap, 393, 101

\bibitem[{{Rayner} {et~al.}(2003){Rayner}, {Toomey}, {Onaka}, {Denault},
  {Stahlberger}, {Vacca}, {Cushing}, \& {Wang}}]{rayner03}
{Rayner}, J.~T., {Toomey}, D.~W., {Onaka}, P.~M., {Denault}, A.~J.,
  {Stahlberger}, W.~E., {Vacca}, W.~D., {Cushing}, M.~C., \& {Wang}, S. 2003,
  \pasp, 115, 362

\bibitem[{{Renzini} \& {Voli}(1981)}]{renzini81}
{Renzini}, A., \& {Voli}, M. 1981, \aap, 94, 175

\bibitem[{{Rose}(1994)}]{rose94}
{Rose}, J.~A. 1994, \aj, 107, 206

\bibitem[{{Rose} {et~al.}(2005){Rose}, {Arimoto}, {Caldwell}, {Schiavon},
  {Vazdekis}, \& {Yamada}}]{rose05}
{Rose}, J.~A., {Arimoto}, N., {Caldwell}, N., {Schiavon}, R.~P., {Vazdekis},
  A., \& {Yamada}, Y. 2005, \aj, 129, 712

\bibitem[{{Serra} \& {Trager}(2007)}]{serra07}
{Serra}, P., \& {Trager}, S.~C. 2007, \mnras, 374, 769

\bibitem[{{Vacca} {et~al.}(2003){Vacca}, {Cushing}, \& {Rayner}}]{vacca03}
{Vacca}, W.~D., {Cushing}, M.~C., \& {Rayner}, J.~T. 2003, \pasp, 115, 389

\bibitem[{{Worthey}(2004)}]{worthey04}
{Worthey}, G. 2004, \aj, 128, 2826

\end{thebibliography}
\end{document}